\documentclass[preprint,a4paper,prc,nofootinbib,showkeys,showpacs]{revtex4-1}
\usepackage{graphicx}

\usepackage[breaklinks]{hyperref}
\hypersetup{colorlinks,urlcolor=black,citecolor=black,linkcolor=black,filecolor=black}
\usepackage{breakurl}

\newcommand{\nucl}[3]{
\ensuremath{
\phantom{\ensuremath{^{#1}_{#2}}}
\llap{\ensuremath{^{#1}}}
\llap{\ensuremath{_{\rule{0pt}{.75em}#2}}}
\mbox{#3}
}}

\def\clap#1{\hbox to 0pt{\hss#1\hss}}

\def\mathrlap{\mathpalette\mathrlapinternal}

\def\mathrlapinternal#1#2{%
\rlap{$\mathsurround=0pt#1{#2}$}}

\begin{document}
\title{OKLO REACTORS AND IMPLICATIONS FOR NUCLEAR SCIENCE}
\author {E. D. DAVIS}
\affiliation{Department of Physics, Kuwait University, P.O. Box 5969 Safat, 13060 Kuwait}
\email{edward.davis@ku.edu.kw}
\author{C. R. GOULD}
\affiliation{Physics Department,  North Carolina State University, 2700 Stinson Drive, 
Raleigh, North Carolina 27695-8202, United States of America\\
Triangle Universities Nuclear  Laboratory, Durham, North Carolina 27708-0308, United States of America}
\email{chris\_gould@ncsu.edu}
\author{E. I. SHARAPOV}
\affiliation{Joint Institute for Nuclear Research, 141980 Dubna, Kurchatov str. 6, Moscow region, Russia}
\email{sharapov@nf.jinr.ru}
\begin{abstract}
We summarize the nuclear physics interests in the Oklo natural nuclear reactors, focusing particularly on developments 
over the past two decades. Modeling of the reactors has become increasingly sophisticated, employing Monte Carlo 
simulations with realistic geometries and materials that can generate both the thermal and epithermal fractions.  The 
water content and the temperatures of the reactors have been  uncertain parameters.  We discuss recent work pointing to 
lower temperatures than earlier assumed. Nuclear cross sections are input to all Oklo modeling and we discuss a 
parameter, the $^{175}$Lu ground state cross section for thermal neutron capture leading to the isomer $^{176\mathrm{m}}$
Lu, that warrants further investigation.  Studies of the time dependence of dimensionless fundamental constants have been 
a driver for much of the recent work on Oklo. We critically review neutron resonance energy shifts and their dependence 
on the fine structure constant $\alpha$ and the  ratio $X_q=m_q/\Lambda$ (where $m_q$
is the average of the $u$ and $d$ current quark masses and $\Lambda$ is the mass scale of quantum chromodynamics).  We 
suggest a formula for the combined sensitivity to $\alpha$ and $X_q$ that exhibits the dependence on proton number $Z$ 
and mass number $A$, potentially allowing  quantum electrodynamic and quantum chromodynamic effects to be disentangled if 
a broader range of isotopic abundance data becomes available. 
\end{abstract}
\pacs{06.20.Jr, 07.05.Tp, 21.10.Sf, 24.30.-v, 27.20.+n, 28.20.Gd, 28.41.-i}
\keywords{Oklo; natural nuclear reactors;  Monte-Carlo simulation; neutron/gamma fluxes; 
core temperature; nuclear data;  nuclear waste depository; time variation fundamental constants.}
\maketitle

\section{Discovery of the Oklo natural  reactors}

On 25 September 1972, Andr\`{e} Giraud, Head of the French {\it Commissariat \`{a} l'\'{E}nergy Atomique} (CEA), 
announced the discovery of a two billion year-old nuclear reactor in Gabon, at the site of the Oklo uranium mines. 
The sequence of events that led to this startling announcement had begun earlier, in June 1972, at the Pierrelatte 
uranium enrichment plant with the observation of a small but definite anomaly in the uranium isotopic ratio for 
a UF$_6$ sample. The supply of anomalous uranium was soon traced to uranium rich ores (up to 60\%) for which 
investigations revealed local uraninite deposits with $^{235}$U isotopic abundance of 0.600\%, instead of the  
normal 0.7202\%.
 
At first glance one can wonder how the Oklo reactors were able to operate when it is well known that the modern 
light water reactors cannot work with natural uranium, requiring instead $^{235}$U
enrichment of about 3.5\%. Natural uranium is composed of three 
isotopes with abundances today\footnote{Abundance and lifetime data are from the 
National Nuclear Data Center (www.nndc.bnl.gov).} of 99.2744\% for $^{238}$U, 
0.7202\% for $^{235}$U, and 0.0054\% for $^{234}$U.
However, the relative enrichment of $^{235}$U increases going back in 
geological time because the half-life of $^{235}$U is 710~Myr
while that of $^{238}$U is 4.51~Gyr. For example,
$^{235}$U enrichment was 1.3\% 700~Myr ago,
2.3\% 1.40~Gyr ago, 4.0\% 2.10~Gyr ago, and 
up to 17\% at the time of creation of the solar system. 

As reviewed 
by Zetterstr\" {o}m \cite{Lena00},
the stabilization of the Oklo area geological basement happened 
not earlier than  2.7 Gyr ago and 
the geological age of
the Francevillian sediments   is estimated \cite{Gau96b}  to be about   
  $2.265\pm 0.15$~Gyr. An independent constraint on the age of the Oklo phenomenon 
is provided by  the Great Oxidation Event \cite{GauW03,Haz09}.  
This happened about 2.2 Gyr ago when, due to the biological activity of
cyanobacteria,  the oxygen content in the
atmosphere of Earth increased by about a factor of a hundred. This allowed uranium to be converted from its insoluble 
uranium(IV) form to its soluble uranium(VI) form.  Deposition of high grade uranium ores in sediments subsequently 
occurred when this soluble uranium was precipitated out, either by reduction back to the insoluble form (by carbon, 
methane or iron), or by direct microbial induced action. 

Interest in natural nuclear reactors preceded their discovery by almost two decades. As early as 1953, 
Wetherhill and Ingram had found evidence in Congo 
pitchblende Xe isotopic data that, besides spontaneous fission, neutron induced fission had taken place \cite{WeIn53}. 
They  stated  that ``the deposit was twenty-five percent of the way to becoming a pile; it is
interesting to extrapolate back 2000 million years \ldots\ Certainly such a deposit 
would be close to being an operating pile.''

Following this suggestion, 
while considering the Johanngeorgenstadt (Saxony) pitchblende -- 
a uranium ore with a minimal content of rare earth poisons,
Kuroda \cite{Kur56} applied Fermi's four-factor pile theory \cite{Fer47} and obtained estimates of neutron multiplication
factors
greater than unity for  proper amounts of water in pitchblende.     
His conclusion was that ``critical uranium chain reactions could 
have taken place if the size of the assemblage was greater than, say,
a thickness of a few feet.'' Kuroda looked for possible changes in the  chemical
composition of samples from uranium ores from several locations (but not Oklo, of course),
and found no signs of {\em chain} reactions.  At the time his paper was written (1956), it seemed highly unlikely 
that natural reactors would be found on Earth. 

The first research reports on Oklo data appeared in the fall of 1972  (Bodu {\em{et al.}} \cite{Bodu72}, 
Neuilly {\em{et al.}} \cite{Neu72} and Baudin {\em{et al.}} \cite{Bau72}) and starting in 1973, the CEA launched 
the project ``Franceville'', named after the town Franceville in the vicinity the Oklo mines. Uranium mining 
was suspended for two years to probe the terrain. Six natural reactor zones were discovered and samples were 
shared widely with the cooperation of  the International 
Atomic Energy Agency (IAEA). In June 1975, the first international Oklo meeting 
took  place in Libreville,   with the proceedings published by the IAEA \cite{IAEA75}.
To continue the work, the IAEA and CEA established an
International Working Group on Natural Reactors with a technical 
committee of experts. This group met in Paris in December 1977 
to review progress, and published further proceedings in 1978 \cite{IAEA78}. A review of 
all work done until 1990 can be found in an excellent book by Naudet \cite{Nau91}, who was the Franceville project head.  

New zones were later identified \cite{Gau96a}, and a European research program ``Oklo - natural
analogue for a radioactive waste repository'' was initiated to study analogies between the
behavior of materials in Oklo and in planned nuclear waste repositories \cite{Bla96}.
This program was financed by the European Commission of the European Union 
and implemented in co-operation with institutions from other countries.

Studies of Oklo continue unabated with about 140 papers in the published literature since 2000. Recent papers split 
evenly between interest in the fascinating geology and operation of the reactors, and interest in what the isotopic 
remains can say about time variation of fundamental constants over the last two billion years.   Notable earlier 
reviews include those by Naudet \cite{Nau76}, Cowan \cite{Cow76}, Petrov \cite{Pet77}, Kuroda \cite{Kur82},
Meshik \cite{Mesh05}, and  Barr\`{e} \cite{Bar05}. We use the basic information from these reviews and  
focus here on recent results on modeling the reactors and their implications for refining bounds on the time 
variation of dimensionless fundamental constants such as the electromagnetic fine structure constant.

\section{Oklo reactors}

\subsection{The fifteen reactor zones}

According to Gauthier-Lafaye \cite{Gau06}, the chief geologist of the project ``Franceville'', fourteen reactor 
zones were located at the Oklo-Okel\'{o}bondo area and one zone at Bangomb\'{e}, 30~km away.  
Other authors \cite{Lae07} cite the total number as seventeen. With regard to the geochemical behavior of the reaction
products, the zones are classified according to whether they were mined close to the surface in open pits (down to about
100~m), or mined underground at greater depth \cite{Lae07}. Zones of the first type (from No. 1 to 9) were certainly 
affected by weathering processes, while zones of the second type were weathered only little, if at all. Of these 
latter zones, reactor zone 10 located at a depth of 400~m  is considered to be  best preserved from
post-reaction alterations.  Located within a ground water discharge area, and  being besides very 
shallow ($10-12$~m deep), the Bangomb\'{e} reactor was in large part washed out. Regrettably, only the Bangomb\'{e}
Site has been preserved \cite{Gau97}. All Oklo-Okel\'{o}{\-}bon{\-}do  zones  were mined out \cite{Gau06}. 

\subsection{Oklo isotopic data}

Prior to excavation, the reactor zones contained gangue, $^{238}$U, depleted $^{235}$U, and stable fission products. 
These products were mostly isotopes of rare earth elements (REEs) resulting from the heavy fission fragments, and some 
elements, such as Zr and Y, resulting from the light fission fragments. As an example, we show in Table 1 the data for Nd,
U, Ce and  Sm taken from Ref.~\cite{Neu72}. Most of the data for Nd and Ce resemble fission yields more closely than 
natural abundance data. The few differences are explained in what follows. 

The isotope $^{142}$Nd is not produced by fission  and therefore its presence in Oklo ores  is due to natural Nd. 
This enables one to make corrections for elements present in the uranium sediments before the chain reaction took place. 
In addition, the elemental abundances and isotopic concentrations in Oklo samples are influenced by neutron capture and 
radioactive decay. Neodymium concentrations are perturbed by the large neutron capture cross sections of $^{143}$Nd and 
$^{145}$Nd. When all the corrections are made, the concentrations of the neodymium isotopes correspond precisely to the 
fission data, and traditionally have provided the best estimates of both the neutron fluence and the total number of 
fissions of  $^{235}$U that occurred in reactor zones.\footnote{Lanthanum isotopic data can, in principle, also give 
estimates of the neutron fluence \cite{Gould4}, but, in practice, such estimates are limited by the lack of isotopic 
data, and are also more susceptible to uncertainties due to the variability in natural elemental abundance.}   
The strong deviation in the Sm data is evidently due to burning of $^{149}$Sm during reactor operation. We will 
discuss the $^{149}$Sm data in detail later; this is the channel of greatest interest in studies of the time 
evolution of fundamental constants.

Since 1972 similar data have been obtained by mass spectrometry methods for many elements. Isotopic composition 
and elemental abundances have been reviewed in Ref.~\cite{Nau91} for reactor zones 2 and 3, in Ref.~\cite{Lae07} 
for reactor zone 9, and in Ref.~\cite{Hid98} for reactor zone 10. In an interesting development, the isotopic 
enrichment of the gaseous fission products Kr and Xe, trapped in minerals, was studied \cite{Shuk77,Mesh00,Mesh04}, 
with important results which we will discuss in the next subsection. One can deduce from the totality of all these 
data the neutron fluence, the average power of the reactors, the duration of reactor operation and the age of the 
Oklo phenomenon (i.e. how long ago it occurred). While some author prefer more recent times of occurrence, the most
accurate determination is reported \cite{Ruf76,Ruf78}  as $1950\pm 40$ Myr, a result in good agreement with the
geological age mentioned above. Results for the duration of reactor operation for Oklo zones 10 and 13 and the 
Bangombe site range from 100 to 300 thousand years \cite{Hid98}.

\begin{table}[pt]
\caption{Isotopic data of the ore sample M from the Oklo mine compared with natural concentrations and with cumulative 
fission yields [Neuilly \emph{et al.}, C. R. Acad. Sci. Paris, Ser. D {\bf 237} (1972) 1847].\footnote{Isotopic
concentrations are in atomic percentages. The relative precision is about 0.2\% for the natural abundances, 
0.5\% for fission  data, and from 2\% to 3\% for Oklo data.} } 
{\begin{tabular}{@{}cccc@{}} 
\toprule
Isotopes  & Natural concentration & $^{235}$U fission & Oklo M\\
&\%  & \% & \%  \\ 
\colrule
$^{142}$Nd\hphantom{00} & \hphantom{0}27.11 & \hphantom{0}0.00 & 1.38 \\
$^{143}$Nd\hphantom{00} & \hphantom{0}12.17 & \hphantom{0}28.8 & 22.1 \\
$^{144}$Nd\hphantom{00} & \hphantom{0}23.85 & \hphantom{0}26.5 & 32.0 \\
$^{145}$Nd\hphantom{00} & \hphantom{0}8.30 & \hphantom{0}18.9 & 17.7 \\
$^{146}$Nd\hphantom{00} & \hphantom{0}17.22 & \hphantom{0}14.4 & 15.6 \\
$^{148}$Nd\hphantom{00} & \hphantom{0}5.73 & \hphantom{0}8.26 & 8.01 \\
$^{150}$Nd\hphantom{00} & \hphantom{0}5.62 & \hphantom{0}3.12 & 3.40 \\
$^{235}$U\hphantom{000} & 0.720 &                      & 0.440\hphantom{0} \\
$^{140}$Ce/$^{142}$Ce\hphantom{00} & \hphantom{0}7.99 & \hphantom{0}1.06 & 1.57 \\
$^{149}$Sm/$^{147}$Sm\hphantom{00} & \hphantom{0}0.92 & \hphantom{0}0.47 & 0.003 \\ 
\botrule
\end{tabular}}
\end{table}

\subsection{Periodic mode of operation}

In most of the earlier work on the Oklo phenomenon, a steady-state mode of operation  was tacitly accepted. With 
this assumption, a neutron fluence in a single zone is of the order of $10^{21}$ cm$^{-2}$ (see, for example, 
Ref.~\cite{Hid98}), and an operation duration of about $3\times 10^{5}$~yr leads to an average neutron flux density 
of $10^{8}$ cm$^{-2}$s$^{-1}$, about five  orders of magnitude less than present day power reactors. 
In a pulsed mode of reactor operation, however, the instantaneous flux can be much higher.

In 1977, Yu. Petrov \cite{Petr77} suggested a self-regulating  mechanism for pulsed operation at Oklo 
based on the negative void temperature coefficient. 
According to this mechanism, with increasing reactor temperature,  
neutron moderating water is converted into steam which can even
leave the active core. As a result, the reactor shuts down, 
cools off, and only restarts when water seeps back 
into the reactor zone again.

The time scale for such operation remained uncertain until 1990 when  Kuroda \cite{Kur90,Kur90-2} came up with the idea 
of estimating the cooling period by analyzing data \cite{Shuk77} on the anomalous isotopic composition of Oklo xenon. 
In particular he compared the excesses (relative to standard cumulative fission yields) of the isotopic ratios 
Xe(132/136) and Xe(134/136) and took note of the fact that 
none of these Xe isotopes is a direct result of fission. Their cumulative yields are due to fission fragment precursors 
having quite different half-lifes. Kuroda explained the excess  
for Xe(132/136) by taking into consideration the  different half-lifes of 
of the precursors, $^{132}$Te (78.2 h) and  $^{134}$Te (0.7 h). 
Assuming  a pulsed mode of operation with a 
reactor-on time, $\Delta\rm{t}_{on}$, 
much larger than the reactor-off time, 
$\Delta\rm{t}_{off}$,  
he found 2.5 h $<\Delta\rm{t}_{off}<$ 3 h. The Xe data of Ref.~\cite{Shuk77}
were for the uranium grains taken directly from several active Oklo cores. 

In the years $2000-2004$, Meshik \emph{et al.} \cite{Mesh00,Mesh04} studied the Xe isotopic anomalies in Oklo zone 13, 
both in uranium grains as well as in Al phosphates formed through 
the action of heated water. They found that Xe was trapped 
preferentially in the phosphates and its concentration exceeded that 
in uranium by several  orders of magnitude. 
The resulting better sensitivity of the isotopic analysis allowed 
them to consider the influence of  many precursors in the mass chains 
leading to the stable Xe isotopes. They were able to deduce values of 
$\Delta\rm{t}_{off}=2.5$ hours and $\Delta\rm{t}_{on}=0.5$ hours. 
Although these results apply strictly only to RZ13, we take the numbers to be typical for all the reactor zones, and use 
them in one of the following sections.

\section{Monte-Carlo simulations of Oklo reactors}

To date, the Oklo uranium deposits are the only known place on Earth where two billion years ago all conditions were
appropriate for fission chain reactions. Those conditions are: (i) a high concentration of uranium in 
different zones, with the $^{235}$U abundance being at least 3\%; (ii) a large amount of water in the pores 
and cracks of minerals, and; (iii) the absence or near-absence of neutron poisons such as boron, lithium and the REEs.  
Elemental compositions and geometries of six Oklo zones 
were already reported at the first Oklo symposium \cite{IAEA75}.
These data were enough to apply theories of neutron chain reactors \cite{Wei58} 
to neutronic calculations for multiplication factors, neutron fluxes 
and other reactor physics parameters. 
Naudet and Filip developed  
the multigroup code BINOCULAR \cite{Naud75}, 
which solved the neutron transport equations for some Oklo zones, especially 
the largest RZ2. Interest in radioactive waste storage lead to the development of sophisticated deterministic codes 
for modeling criticality and these have also been applied to Oklo \cite{Mas011,Mas012}. 

Most recently, a probabilistic approach 
based on the application of the code MCNP \cite{Brie00} has been utilized. The MCNP code 
models neutron transport using Monte Carlo methods and takes advantage of 
continuous libraries of energy dependent neutron cross sections.  
It can deal with practically any realistic geometry of an active reactor core. Modeling with MNCP has been 
performed for zones RZ2 \cite{Petr06,Gould1},    
RZ3 \cite{Oneg012},  RZ9 \cite{Salah011,Ben011}, and RZ10 \cite{Gould1}. We illustrate some details of such 
modeling using the examples of zones RZ2 and RZ10.

\subsection{Active core composition and criticality}

The MCNP code input for any Oklo reactor consists of a description of its geometry with relevant physical dimensions, 
the operating temperature, and specification of the amounts of uraninite 
(a variety of UO$_2$), gangue and water. Typically, reactor zones 
are lens shaped layers, about 10 m long, about 10 m wide and up to 0.8 m thick.
The uraninite content varies from zone to zone reaching as  high as 80wt.\% in some samples.  
The gangue is composed of clays -- the hydrated silica-aluminates 
of different metals. Data for several zones are detailed in Ref.~\cite{Nau91}.
The amount of water during the time of reactor operation 
remains very uncertain. Water, however, defines the shape of the neutron 
spectrum and can therefore be deduced by varying its amount until the 
spectral indices\footnote{Spectral indices are a measure of the epithermal neutron fraction in the neutron 
spectrum.} of the calculated neutron flux coincide with the indices deduced from the isotopic data. 
The output of MCNP consists of the criticality value 
and the energy dependence of the neutron spectrum.

In Ref.~\cite{Gould1}, the geometries of Oklo zones RZ2 and RZ10 were approximated 
as flat cylinders of 6 m diameter and 70 cm height,  surrounded by a 1 m thick reflector
consisting of water saturated sandstone. The compositions (in g/cm$^3$) of the two reactors
are shown in Table 2. The composition of RZ2 is from Ref.~\cite{Nau91} and that of RZ10 
from Refs.~\cite{Gau96a} and \cite{Hid98}. The total density of the active core material 
at ancient times was about 4~g/cm$^3$ for RZ2 and 3~g/cm$^3$ for RZ10.
The most striking difference between RZ2 and RZ10
is the rather small, 28wt.\% content of uraninite UO$_2$ in RZ10.  As a result, 
RZ10 cannot become critical with a poison more than 1~ppm of $^{10}$B equivalent 
while RZ2 can accommodate 10~ppm of $^{10}$B equivalent.
The amount of water (H$_2$O) shown in Table 2 is the total amount, including water of
crystallization.

\setlength{\tabcolsep}{5pt}

\begin{table}[pt]
\caption{Composition (in g/cm$^3$), multiplication coefficient k$_\mathit{eff}$ and spectral index r$_O$  
of  Oklo reactors [C. R.  Gould, E. I.  Sharapov, S. K. Lamoreaux, Phys. Rev. C {\bf 74} (2006) 24607]. }
{\begin{tabular}{@{}cccccccccccc@{}} \toprule
Zone  & UO$_2$ & H$_2$O & SiO$_2$ & FeO & 
Al$_2$O$_3$ & MgO  & MnO & K$_2$ & Total & k$_\mathit{eff}$\footnote{The k$_\mathit{eff}$ values correspond to 
poisoning equivalent to 10 and 0.8 ppm of $^{10}$B  for zones RZ2 and RZ10, respectively.}
& $r_O$  \\ 
\colrule
RZ2 &2.500 &0.636 &0.359 &0.149 &0.238 &0.077 &0.009 &0.020 &3.99 &1.033 
&0.22 \\
RZ10 &0.850 &0.355 &0.760 &0.320 &0.510 &0.160 &0.020 &0.040 &2.96 &1.036 
& 0.15 \\ 
\botrule
\end{tabular}}
\end{table}

\setlength{\tabcolsep}{6pt}

\subsection{Neutron and $\gamma$-ray fluxes}

The MCNP code allows one to model both the neutron fluxes in reactors and $\gamma$-ray fluxes. 

\begin{figure}[tp]
\centering
\includegraphics[width=8.6cm]{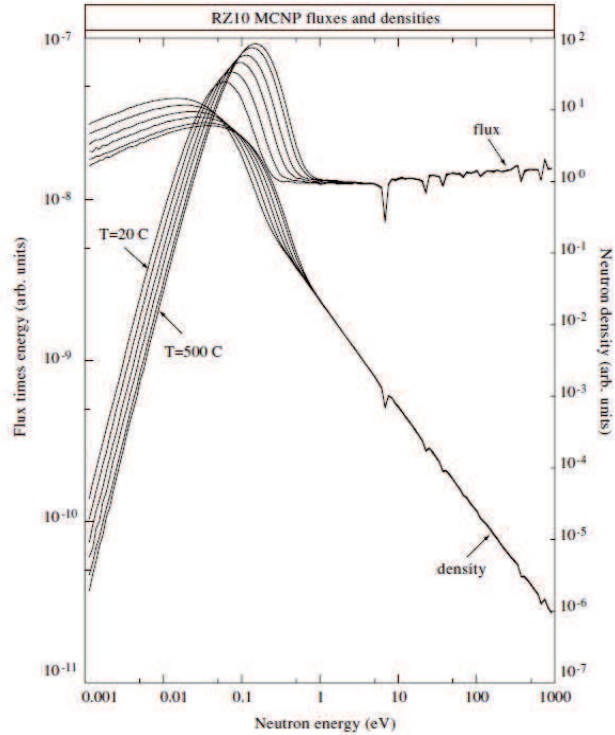}
\caption{Reactor zone RZ10 MCNP neutron fluxes and neutron densities for
different temperatures [C. R.  Gould, E. I.  Sharapov, and S. K. Lamoreaux
Phys. Rev. C {\bf 74} (2006) 24607]. The neutron fluxes are
plotted as $E\,\phi(E)$ and are the family of curves starting from a temperature
of 20~$^\circ$C at the lower left. The neutron densities
are the families of curves starting upper left.}
\label{fig:rz10flux}
\end{figure}

Because the temperature in the active Oklo cores remains uncertain (see later discussion), 
Ref.~\cite{Gould1}  obtained the neutron fluxes of zones RZ2 and RZ10  for a set of different 
temperatures: 20~$^\circ$C, 100~$^\circ$C, 200~$^\circ$C, 300~$^\circ$C, 400~$^\circ$C, 
and 500~$^\circ$C. The calculations of the energy dependent neutron fluxes were performed 
with energy bins  on a
lethargy grid, where the (dimensionless) lethargy $u$ ran from 23 to 9.3 in steps of 0.1. This gave 
neutron energies $E = (10^7\,\mathrm{eV}) e^{-u}$ from about 1 meV up to about 1 keV,
and bin widths $\Delta E \approx 0.1 E$. In Fig.~\ref{fig:rz10flux},
we show the modeled spectra as a function of neutron energy $E$ for RZ10. 
The neutron fluxes are the family of curves starting at
the \emph{lower\/} left of the figure.
The leftmost curve corresponds to a temperature of 20~$^\circ$C and the rightmost curve to 500~$^\circ$C. Uranium 
absorption resonances are prominent in the epithermal region. 

For some purposes, the neutron density is a more relevant than the neutron flux. The neutron flux $\phi (E)$ is 
related to the neutron density $n(E)$ by $\phi(E) = n(E)v(E)$, where $v(E)$ is neutron velocity. Neutron densities
(normalized to one neutron per unit volume) are the families of curves starting \emph{upper\/} left in 
Fig.~\ref{fig:rz10flux}. The topmost curve corresponds to 20~$^\circ$C and the lowest curve corresponds to 
500~$^\circ$C. These spectra demonstrate clearly the presence of two components in two different energy regions: 
the thermal neutron region below about $0.5-1.0$~eV and the so-called epithermal or $1/E$ region.

\begin{figure}[th]
\centering
\includegraphics[width=6.3cm]{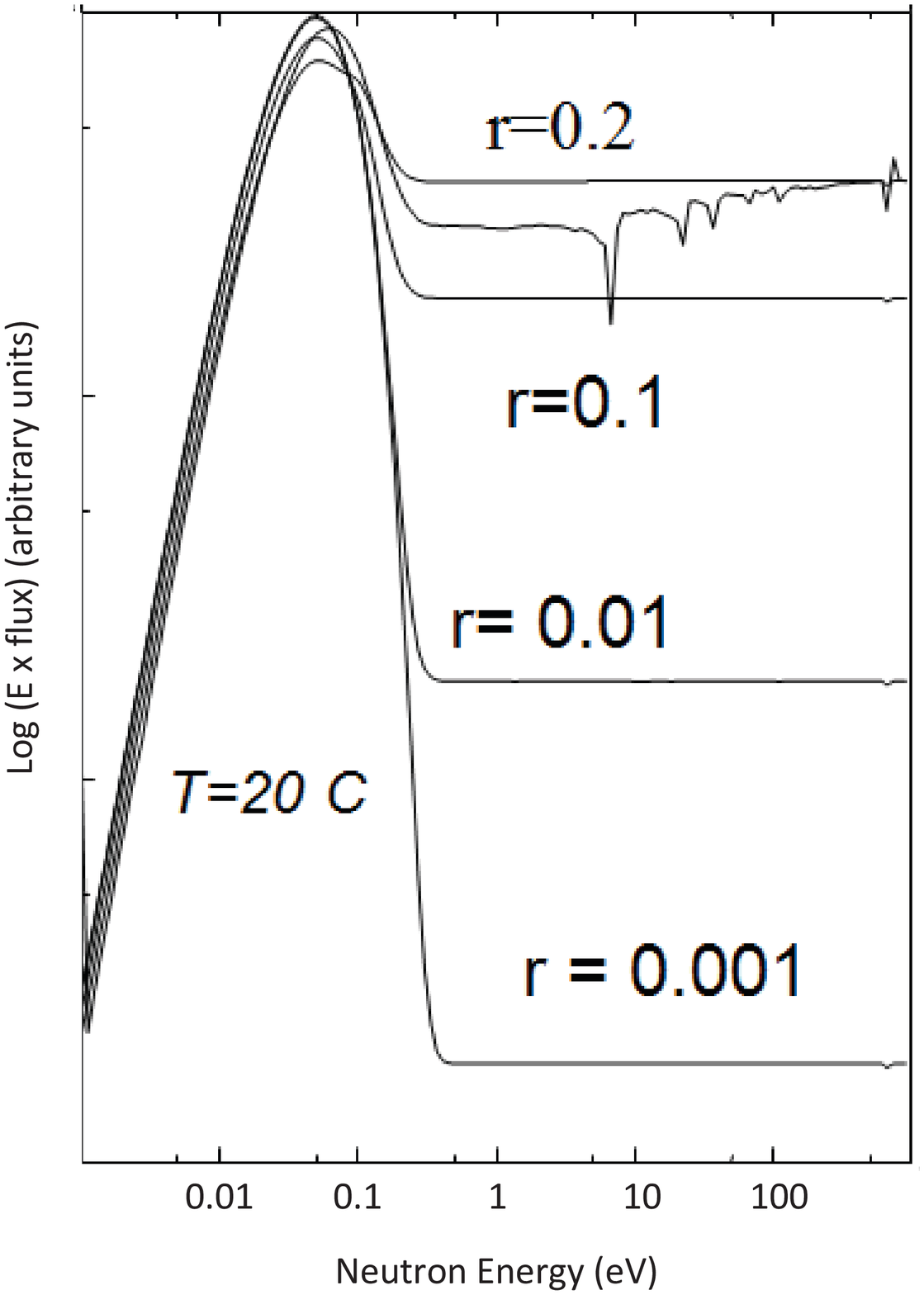}
\caption{Realistic RZ10 MCNP neutron flux at thermal energies compared to  model fluxes with different Oklo 
spectral indices $r_O$.}
\label{fig:rz10fluxcompare}
\end{figure}

Despite the fact that the MCNP spectra do not exactly follow $1/E$-behaviour in the epithermal region, 
Oklo spectral indices $r_O$ can be deduced from flux plots. Figure \ref{fig:rz10fluxcompare} illustrates the 
methodology and values determined from Ref.~\cite{Gould1} for $r_O$ are listed in
Table 2. They agree with ``measured'' values deduced from analysis of concentrations of $^{235}$U and of the 
fission products $^{143}$Nd and  $^{147}$Sm \cite{Nau91,Gau96a,Hid98,Ruf76}:
$r_O=0.20 - 0.25$ for RZ2 and $r_{O}=0.15 \pm 0.02$ for RZ10.
   
\begin{figure}[tp]
\centering
\includegraphics[width=8.6cm]{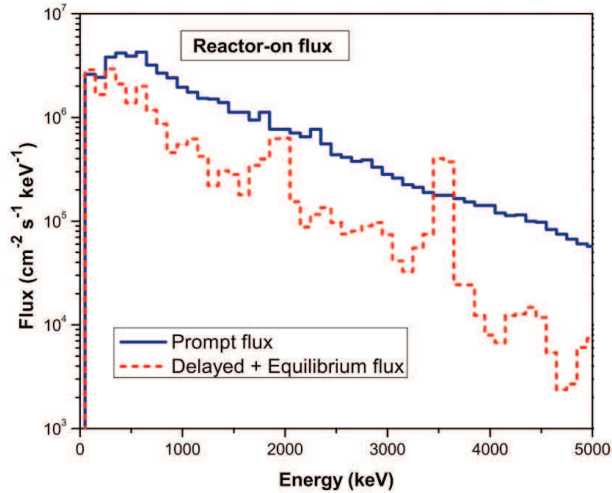}
\caption{Prompt and delayed $\gamma$-ray fluxes in Oklo zone RZ10. The fluxes are calculated for an 18~kW reactor, 
on for 0.5 hr and off for 2.5 hr. The prompt flux is the upper line. The lower line is the delayed flux, taking into
account the equilibrium flux  associated with N previous reactor-on pulses ($\mathrm{N}>>1$).}
\label{fig:gflux}
\end{figure}

Gamma-ray fluxes can, in principle, contribute to post-processing of fission products. The issue of whether this was a 
concern for Oklo was investigated for the first time in Ref.~\cite{Gould3}. Gamma-rays in a reactor arise in three ways:
\begin{itemize} 
\item prompt $\gamma$-ray emission during fission events 
(eight $\gamma$-rays per fission, with a total energy of about 
7.0 MeV), 
\item prompt $\gamma$-ray emission following neutron radiative capture in  materials of the reactor (also with a
total energy of about 7.0 MeV) 
 \item delayed $\gamma$-ray emission from 
the decay of fission products (a total energy of about 6.3 MeV). 
\end{itemize}
Prompt $\gamma$-production was modeled for the Oklo reactor zone RZ10 with the MCNP code using the same input files 
as for the neutron flux modeling. The modeling was performed for the pulsed mode of the reactor operation discussed 
above (reactor on for 0.5~h and off for 2.5~h). Delayed $\gamma$-production as a function of time was studied    
analytically by Way \cite{Way48} and later numerically with improved decay data libraries \cite{algora10}. 
The results of Ref.~\cite{Gould3} are shown in Fig.~\ref{fig:gflux}. As might be expected, the total prompt 
$\gamma$-ray flux dominates and is about $3\times 10^9$ $\gamma$ cm$^{-2}$ s$^{-1}$. However, at some energies, 
the delayed $\gamma$-ray flux is significant. Implications of these calculations for lutetium thermometry 
in RZ10 will be discussed in the next section.

\section{Reactor temperatures}

Much attention has been focused on the temperatures at which the Oklo reactors operated. This parameter sets the shape 
of the energy spectrum, an important input to Oklo analyses investigating possible time variations in fundamental
constants. The pulsed mode of operation currently favored indicates no single temperature can be appropriate.
Nevertheless an average temperature is crucial to establishing general characteristics, and will be an appropriate
parameter as long as the system is responding in a linear fashion. 

The reactors likely operated under conditions similar to present day PWR systems, with pressures about 150~atm 
and temperatures of about 300~${}^\circ\mathrm{C}$. The critical point for water (no liquid possible at any 
pressure) occurs at about 374~${}^\circ\mathrm{C}$ and 218~atm. This sets a plausible upper limit to the 
temperature although some modeling of Oklo reactors has achieved criticality at higher temperatures, even up to 
about 550~${}^\circ\mathrm{C}$ in some cases \cite{Petr06}.
    
Lutetium thermometry first proposed by Wescott \cite{West58} can in principle provide a direct measure. The method
depends on the overlap of the thermal neutron spectrum with a low energy neutron capture resonance, in this case at 
141~meV in $^{176}$Lu. The temperature is extracted by comparing the $^{176}$Lu/$^{175}$Lu ratio and noting the 
depletion of $^{176}$Lu will be strongly dependent on temperature whereas the depletion of $^{175}$Lu is not. 
Effective cross sections for the two isotopes are listed in Table~\ref{tb:Lucs}.
  
\begin{table}[pt]
\caption{Lutetium effective cross sections $\hat\sigma_{5}$ and 
$\hat\sigma_{6}$ for the Oklo RZ10 reactor at temperatures 
$T_O$ from 0 $^\circ$C to 600 $^\circ$C. (C. R.  Gould and E. I.  Sharapov,
    Phys. Rev. C {\bf 85} (2012) 024610). \label{tb:Lucs}}
{\begin{tabular}{@{}ccc@{}} \toprule
$T_O$ ($^\circ$C) &$\hat\sigma_{5}$ (kb) &$\hat\sigma_{6}$ (kb)\\ 
\colrule
0 &  0.115 & 4.216 \\
20 & 0.115 & 4.487 \\
100 & 0.115 & 5.359 \\
200 & 0.115 & 6.310 \\
300 & 0.114 & 7.013 \\
400 & 0.114 & 7.544 \\
500 & 0.114 & 7.715 \\
600 & 0.114 & 7.750 \\
\botrule
\end{tabular}}
\end{table}

The method was first applied to Oklo studies by Holliger and Devillers \cite{Holli81} who found temperatures of 
$T_O =  260~{}^\circ\mathrm{C}$ and 280~${}^\circ\mathrm{C}$ for RZ2 and RZ3, respectively. The later more realistic
modeling of Onegin \cite{Oneg012} found $T_O = (182 \pm 80)~{}^\circ\mathrm{C}$ for RZ3. However for RZ10, one of 
the most well characterized zones, Hidaka and Holliger \cite{Hid98} succeeded  in getting a result 
($T_O =380~{}^\circ\mathrm{C}$) for only one sample. Three other samples yielded only an anomalously high bound, 
$T_O > 1000~{}^\circ\mathrm{C}$. 

Reasons for these problematic results were investigated by Gould and Sharapov \cite{Gould2}. They noted the dependence 
on a potentially uncertain parameter, $B^g$, the $^{175}$Lu ground state to isomer state capture branching ratio. 
This issue emphasizes anew the role of nuclear data in understanding the Oklo phenomenon, a point which was stressed 
already \cite{Rho80} at the beginning of studies of Oklo. Since then most of the nuclear data required for Oklo analyses 
has been improved considerably, in particular the mass distributions of the fission products of $^{235}$U, 
$^{238}$U and $^{239}$Pu, the isotopic ratio data  Nd/$^{238}$U and Bi/$^{238}$U, 
the Sm, Gd and Nd  isotopic ratios, and various decay constants.  
Nevertheless, it appears that the lutetium branching ratio $B^g$ for thermal neutrons remains very uncertain.
This ratio has been the subject of intensive study by the astrophysics community because of its importance in 
understanding the s-process in stars. New measurements \cite{Klay91,Wiss06} at 5 and 25 keV gave values for $B^g$ 
significantly different from the thermal energy value which is implied by
the neutron cross sections in Mughabghab's \emph{Atlas of Neutron Resonances} \cite{Mug06} and needed for thermometry.
Using the most accurate of these improved ``astrophysical'' values, the meta-sample reanalysis for RZ10 yields $T_O = 
(100 \pm 30)~{}^\circ\mathrm{C}$, much lower than the earlier values cited. The Onegin RZ3 result will also change 
if a lower value of $B^g$ is used, being instead closer to about 100~$^\circ\mathrm{C}$. 
However, because of the presence  of p-wave neutrons at keV energies
(which are absent at thermal energies), the $B^g$ values of Refs.~\cite{Klay91} and \cite{Wiss06}
may not be directly applicable to thermal neutrons.
A precision measurement of $B^g$ at thermal energies would clearly help to resolve this issue
and ways of performing this experiment have been pointed out in Ref.~\cite{Sha013}.

An alternate explanation for problems with lutetium thermometry, burn up of $^{176}$Lu in the $\gamma$-ray flux of 
the reactors, was investigated in Ref.~\cite{Gould3}. Again, improved nuclear data is required, this time for the
integral cross sections \cite{Thrane,Mohr09} for photo-excitation of the specific excited states of $^{176}$Lu which 
decay to metastable $^{176\mathrm{m}}$Lu. Excitation of this isomeric state (which then beta decays to $^{176}$Hf) is 
an important loss mechanism in astrophysical environments \cite{Carr91}. 
However, the $\gamma$-ray fluxes in Oklo reactors obtained 
in Ref.~\cite{Gould3}  prove to be many orders of magnitude too low for 
these ($\gamma,\gamma^\prime$)-reactions
to impact application of lutetium thermometry. 
   
In summary, despite uncertainties, most of the later Oklo analyses together with the latest nuclear data favor 
lower temperatures compared to earlier work.

\section{Geological repository of nuclear wastes}

Oklo is unique not only for being the site of multiple natural reactors, but also as a place where fission 
products have been preserved undisturbed for over two billion years. The zones therefore present 
a unique opportunity for learning more about the long-term geochemical behavior
of radioactive wastes from nuclear reactors. This was realized from the very
beginning by US scientists who accordingly started an Oklo program
relevant to waste management technology \cite{Cow78}.

The chief concern of this program was determination 
of the relative mobilities of the reactor products in the 
Oklo environment and the migration/retention 
behavior of various elements. One of the most important, and surprising, early   
findings was that uranium and most of the REEs did not experience 
significant mobilization in thepast two billion years. Results have been published in 
proceedings of conferences \cite{IAEA75,IAEA78} and reviewed in some detail in Ref.~\cite{Cow78}. 
Further work followed in a European Commission Program ``Oklo -- natural analogue for a
radioactive waste repository''. The focus here was on the considerably more complicated situation involving storage 
of waste materials from modern high power reactors. Results obtained from this program can be found in 
Refs.~\cite{Bla96,Berz90} and \cite{Gau02}. Modeling the potential for criticality in geological storage is an 
important issue and continues to be studied intensively.  
Here again the Oklo reactor zones have served as a testing ground 
for new approaches and new computation codes \cite{Recha03,Mas011,Mas012,Pad012}. 

Because the wastes were contained successfully in Oklo, it appears not unrealistic to hope that  
long term disposal in specially selected and engineered 
geological repositories can be successful \cite{Ewi02}.

\section{Stability of the fine structure constant and light quark masses}

The late Alexander Shlyakhter was the first to recognize the potential of Oklo neutron absorption cross sections
to constrain fluctuations in fundamental constants \cite{Shl76}. He seems to have derived his inspiration from an 
appealing analogy with the ordinary radio receiver (to which he alludes): just as the reception frequency of a 
radio is altered when the parameters of its tuning circuitry are adjusted so the energy of a neutron capture 
resonance shifts as ``constants'' in the nuclear Hamiltonian change. Armed with the confidence of a pioneer
who knows there is no earlier explorer to contradict him, Shlyakhter presented crude order of magnitude estimates
to demonstrate the sensitivity of resonance shifts (expounded with a few short sentences). He deduced upper bounds on the 
time-dependent variation of the strong, weak and electromagnetic coupling constants, which were 
significantly better than previous upper bounds due to Dyson \cite{Dys67,Dys72} and Davies \cite{Davies}.

The intervening years have witnessed a number of attempts to make more convincing estimates with the Oklo energy shift 
data \cite{Irvi83,SV90,SV91,Dam96,FIFO00,FIFO02,FS02,OPQC02,DF03,FS03,Fuj04,Lam04,Flam09}.
Some of this work has been driven by the realization that, within many theories beyond the standard 
models of particle physics and cosmology, fundamental constants are, in effect, dynamical 
variables \cite{Mar84,DP94} (see also the reviews in Refs.~\cite{Uza03} and \cite{Uza2011}). More pressing,
perhaps, have been the empirical findings \cite{Web01,MFWD04,Sri04,Pei04,Webb11} suggesting a change in the fine structure 
constant $\alpha$ over cosmological time scales. As emphasized in Refs.~\cite{Lan02} and \cite{Cal02}, grand unification schemes 
automatically imply that other fundamental Standard Model parameters should evolve with time. 
Relating the Oklo phenomenon to the behaviour of parameters in quantum chromodynamics (QCD) has been frustrated
by the difficulties of the confinement problem compounded by the usual uncertainties of the nuclear many-body 
problem. An elegant scheme for extracting information on the fine structure constant $\alpha$ of quantum electrodynamics (QED) has, however, been developed \cite{Dam96}.
In Refs.~\cite{SV90} and \cite{Dam96}, the implications of Oklo for the weak interaction (specifically, the 
Fermi coupling constant $G_F$) are considered briefly, but it would seem reasonable to ignore them (as 
other authors have done): in the absence of parity-violation, electromagnetic and QCD effects mask any 
contribution of the weak interaction to neutron capture. 

Despite the appeal in Ref.~\cite{Lan02} to obtain 
``more accurately the constraints that [the Oklo reactor] imposes on the space of coupling constants'', most 
analyses of Oklo restrict themselves to inferring the corresponding limits on the change in $\alpha$ only.
The $97.3\,\mathrm{meV}$ resonance seen in neutron capture by ${}^{149}$Sm (which we term \emph{the\/} Sm 
resonance below) has received most attention.

\subsection{Sensitivity of an energy shift to the fine structure constant} \label{sc:alpha}

The analysis of Damour and Dyson in Ref.~\cite{Dam96} begins with the approximation of the neutron capture
resonance energy $E_r$ as a difference in expectation values of the nuclear Hamiltonian $\widehat{H}$:
\begin{equation}\label{er}
 E_r = \langle r| \widehat{H} | r\rangle - \langle g | \widehat{H} | g\rangle,
\end{equation}
where $|g\rangle$ denotes the eigenket of the ground state of the target nucleus and $|r\rangle$ denotes the eigenket
of the compound nucleus state formed by the capture of the neutron. By the Feynman-Hellmann
theorem, the rate of change of $E_r$ with $\alpha$ is
\[
 \frac{d E_r}{d\alpha} = \langle r| \frac{\partial \widehat{H}}{\partial \alpha} | r\rangle 
                            - \langle g | \frac{\partial \widehat{H}}{\partial \alpha} | g\rangle .
\]
If, as in Ref.~\cite{Dam96}, small electromagnetic effects (like magnetic-moment interactions and QED radiative 
corrections to nucleon properties) are neglected, then only the Coulomb potential energy operator $\widehat{V}_C$ 
(proportional to $\alpha$) contributes to the partial derivative of $\widehat{H}$ --- i.e.
\begin{equation}\label{dVc}
 \alpha \frac{d E_r}{d\alpha} = \langle r| \widehat{V}_C | r\rangle 
                            - \langle g | \widehat{V}_C | g\rangle .
\end{equation}
To evaluate the difference in Coulomb energies in Eq.~(\ref{dVc}), Damour and Dyson resort to 
replacing the expectation values by the classical Coulomb energies
\[
      \textstyle{\frac{1}{2}} \int V_i \rho_i\, d^3r,
\]
where $V_i$ is the electrostatic potential associated with the charge density $\rho_i$. With this 
approximation, Eq.~(\ref{dVc}) becomes
\begin{equation}\label{dVcc}
  \alpha \frac{d E_r}{d\alpha} \simeq {\textstyle\frac{1}{2}} \int\! ( V_r \rho_r - V_g \rho_g )\, d^3r .
\end{equation}
The ingenious step of Damour and Dyson is to now use Poisson's equation along with Green's second identity and 
recast the right-hand side of Eq.~(\ref{dVcc}) in the form
\begin{equation}\label{adVcc}
   \int\! V_r \delta \rho\, d^3r - {\textstyle\frac{1}{2}} \int\! \delta V \delta \rho\, d^3r
\end{equation}
in which $\delta \rho = \rho_r - \rho_g $ and $\delta V = V_r - V_g$. 

Equation~(\ref{adVcc}) permits two simplifications. The integral in the second term is positive because it is 
twice an electrostatic self-energy. For the purposes of establishing an upper bound on $dE_r/d\alpha$, the 
second term in Eq.~(\ref{adVcc}) can be discarded. The second more subtle advantage of Eq.~(\ref{adVcc}) is 
that one can exploit the fact that any effects of deformation (either static or dynamic) should be less 
pertinent to the excited compound nucleus state $|r\rangle$ than they are to the ground state
$|g\rangle$. 
Damour and Dyson adopt for $V_r$ the potential $V_u$ of a uniformly charged sphere of radius $R_r$ and charge 
$Q=Ze$ which they claim implies that
\begin{equation}\label{intVr}
 \int\! V_r \delta \rho\, d^3r \approx \int\! V_u \delta \rho\, d^3r
                                       = - \frac{(Ze)^2}{2R_r^3}\,  \left[r^2\right]_{rg}, 
\end{equation}
where
\[
 \left[ r^2 \right]_{ij} \equiv \frac{1}{Ze} \int\! r^2\,(\rho_i-\rho_j)\, d^3r .
\]
Determination of an upper bound on $d E_r/d\alpha$ is reduced to finding a positive lower bound on
$\left[ r^2 \right]_{rg}$.

To this end, it suffices to make the reasonable supposition that the charge distribution
is more diffuse in the excited state $|r\rangle$ than in the ground state $|\widetilde{g}\,\rangle$ 
of the same nucleus or that
$\left[ r^2 \right]_{r\widetilde{g}} \ge 0$. It follows that
\[
 \left[ r^2 \right]_{rg} = \left[ r^2 \right]_{r\widetilde{g}} + \left[ r^2 \right]_{\widetilde{g}g}
                        \ge \left[ r^2 \right]_{\widetilde{g}g} \, ,
\]
where, significantly, the (positive) numerical value of $\left[ r^2 \right]_{\widetilde{g}g}$ is obtained 
directly from 
experimental isotope-shifts \cite{BSS80} (see Ref.~\cite{Dam96} for details). 
Finally, one obtains the inequality
\begin{equation}\label{DDresult}
  \alpha \frac{dE_r}{d\alpha} < - \frac{(Ze)^2}{2R_r^3}  \left[ r^2 \right]_{\widetilde{g}g} ,
\end{equation}
which is the result Damour and Dyson use in their interpretation of Oklo data.

The analysis of Damour and Dyson 
leading to Eq.~(\ref{DDresult})
has been subject to some criticisms, most trenchantly in Ref.~\cite{FIFO00}.
Remarkably, there is one implicit and unfounded assumption that seems 
to have gone unnoticed. There are also a number of nuclear structure issues which are not addressed in 
Ref.~\cite{Dam96} or elsewhere. We discuss these matters, before taking up the comments of 
Ref.~\cite{FIFO00}.

In their derivation of Eq.~(\ref{intVr}), Damour and Dyson take the potential $V_u$ of the uniformly charged 
sphere to be
\[
   V_u=V^\mathrm{inside} \equiv \frac{Ze}{2R_r}\Biggl[ 3 - \left(\frac{r}{R_r}\right)^2 \Biggr]
\]
for \emph{all\/} radial distances $r$ (from the center of the charge distribution), whereas, in fact, 
$V_u=V^\mathrm{inside}$ only for $r<R_r$, while, for $r\ge R_r$, 
$V_u = V^\mathrm{outside} \equiv Ze/r$. Accordingly, Eq.~(\ref{intVr}) should, in principle, be replaced by the 
far less appealing result
\begin{equation}\label{myVrint}
 \int\! V_r \delta \rho\, d^3r \approx - \frac{(Ze)^2}{2R_r^3}\,  \left[r^2\right]_{rg} \ +\
            \int\limits_{R_r}^\infty\! \left( V^\mathrm{outside} - V^\mathrm{inside}\right)\! \delta\rho\, d^3r .
\end{equation}
A back of envelope estimate suggests that the integral on the right-hand side of Eq.~(\ref{myVrint}) is 
com{\-}parable to the first term (the contribution retained by Damour and Dyson). The sign of the integrand 
in this integral is determined by the sign of $\delta \rho$, which is almost certainly positive for 
$r\ge R_r$ (in view of the greater diffusiveness of the charge distribution of $|r\rangle$). As seen
above, the first term on the right-hand side of Eq.~(\ref{myVrint}) is negative. Thus, unfortunately, the 
possibility exists that there could be a significant cancellation.

The approximation of $V_u$ by $V^\mathrm{inside}$ for all values of $r$ is in the spirit of early 
studies \cite{La80} of the effect of the Coulomb interaction on nuclear properties, where the rapid drop off 
in nuclear wave functions was presumed to guarantee that the difference between $V_u$ and $V^\mathrm{inside}$ 
for large $r$ has negligible effect. It is now known that, for $Z\ge 28$, the electric field outside 
the nucleus contributes typically about 80\% of a nuclear Coulomb energy \cite{Jae72}. 

The earlier
simplification leading to Eq.~(\ref{dVcc}) amounts to retaining only, in the parlance of nuclear many-body 
theory, the \emph{direct\/} part of the difference in Coulomb energies. One can seek to improve upon this, in 
principle, by including an exchange contribution (arising from the fermion character of nucleons) and a spin-orbit 
term (proportional to the derivative of the electrostatic potential). The importance of such corrections 
can most simply be gauged with the analytic expressions for Coulomb energies of section 9 in 
Ref.~\cite{Jae72}. These formulae have been inferred from a phenomenological study of Coulomb displacement 
energies for spherical nuclei with $Z\ge 28$ and contain multiplicative corrections accommodating the 
diffuseness of the charge distribution's surface. A prescription for estimating the effect of deformation 
is also given, namely to reduce the direct Coulomb energy by a factor of $(1-\beta^2/4\pi)$, where 
$\beta$ is the quadrupole deformation parameter. In fact, the results of Ref.~\cite{Jae72} imply that, 
for the isotopes considered in Ref.~\cite{Dam96} (and subsequent analyses of Oklo data), all the above-mentioned effects induce a decrease of less than 3\% in a ground
state Coulomb energy. Hence, Damour and Dyson are completely justified in neglecting them.

As regards corrections due to inter-nucleon correlations (a topic not considered explicitly in 
Ref.~\cite{Jae72}), it can be argued that these should be negligible in an analysis based on
Eq.~(\ref{myVrint}). For the purposes of establishing an upper bound, $[r^2]_{rg}$ is replaced by 
$[r^2]_{\widetilde{g}g}$, which is taken from experiment and, hence, automatically includes any such
corrections. Since the integral in Eq.~(\ref{myVrint}) is over a region of low nucleon density, it
should be insensitive to short range correlations. Long range correlations are responsible for nuclear
deformation, the effect of which on Coulomb energies has already been discounted.

We can summarize our analysis as follows: for the purpose of order of magnitude estimates, it is 
plausible that
\begin{equation}\label{ome}
   \left| \alpha \frac{dE_r}{d\alpha} \right| \sim \frac{(Ze)^2}{2R_r^3}\,  \left[r^2\right]_{\widetilde{g}g},
\end{equation}
where $R_r$ is the equivalent charge radius of the neutron capture resonance
and $[r^2]_{\widetilde{g}g}$ is the difference between the mean square charge
radii of the ground states $|\widetilde{g}\rangle$ and $|g\rangle$ (of the daughter and target nucleus, 
respectively). We have dropped the integral in Eq.~(\ref{myVrint}) to recover what looks like the result 
of Damour and Dyson, but there are two differences. First, we do not believe that it is appropriate to associate
confidence levels with bounds extracted using Eq.~(\ref{ome}). Second, we have not committed ourselves to the
numerical values of $R_r$ and $[r^2]_{\widetilde{g}g}$ adopted in Ref.~\cite{Dam96} for the
$E_r=97.3\,\mathrm{meV}$ resonance in neutron capture on $^{149}$Sm.

To close, we address briefly the concerns expressed in Ref.~\cite{FIFO00}. The validity of Eq.~(\ref{dVc})
is questioned on the grounds that the nuclear kinetic energy depends implicitly in $\alpha$
(through the role of Coulomb repulsion in determining the size of a nucleus). Our presentation 
makes clear that Eq.~(\ref{dVc}) is a straightforward consequence of the Feynman-Hellmann theorem.
For the derivative of a stationary state expectation value with respect to a parameter in the Hamiltonian, only 
\emph{explicit\/} dependence on the parameter is relevant, not any implicit dependence
of the stationary state eigenfunctions. The Feynman-Hellmann theorem is, admittedly, counter-intuitive.

The other issue raised in Ref.~\cite{FIFO00} is the need to extend the analysis of Damour and Dyson
to include the time dependence of parameters of the strong nuclear interaction $\widehat{V}_N$. The dependence 
on a (dimensionless) strength $\alpha_s$ of $\widehat{V}_N$ is considered. The focus on $\alpha_s$ is 
unfortunate because its relation to parameters of QCD is unclear (as the authors of Ref.~\cite{FIFO00} 
readily acknowledge). The subsequent analysis is invalidated by the omission of the contribution of the kinetic 
energy operator $\widehat{T}$ to the relation for $E_r$ in terms of $\alpha_s$ and $\alpha$. 
From Eq.~(\ref{er}), with \cite{HF07} $\langle\langle\widehat{A} \rangle\rangle\equiv\langle r|\widehat{A} 
|r\rangle - \langle g|\widehat{A} |g\rangle$,
\[
  E_r = \langle\langle \widehat{T} \rangle\rangle + \langle\langle \widehat{V}_N \rangle\rangle
                                                  + \langle\langle \widehat{V}_C \rangle\rangle ,
\]
whereas Ref.~\cite{FIFO00} employs (in our notation) 
$E_r=\langle\langle \widehat{V}_N\rangle\rangle +\langle\langle \widehat{V}_C\rangle\rangle$.
A Fermi gas model estimate of $\langle\langle \widehat{T}\rangle\rangle$ for the Sm resonance considered by Damour and Dyson is $23.5\,\mathrm{MeV}$,
an order of magnitude bigger than $\langle\langle \widehat{V}_C\rangle\rangle$! (From above, 
$\langle\langle \widehat{V}_C\rangle\rangle\sim -1\,\mathrm{MeV}$.)

\subsection{Sensitivity of an energy shift to light quark masses}\label{sc:sqm}

In Ref.~\cite{Dam96}, it was recognized (by appealing to the character of chiral perturbation theory) that 
Oklo data could, in principle, be used to place constraints on the time variation of the mass ratios $m_l/m_p$, 
where $m_l$ denotes the mass of either light (u or d) quark and $m_p$ is the proton mass. 
Despite the fact that there are good reasons \cite{Wil07} for using $m_p$ as the mass unit for problems involving QCD 
and nuclear physics in preference to the mass scale $\Lambda$ of QCD, 
work subsequent to Ref.~\cite{Dam96} has considered the dimensionless ratio 
$X_q=m_q/\Lambda$, where $m_q=\frac{1}{2}(m_u+m_d)$. The most complete analysis to date is that of 
Flambaum and Wiringa in Ref.~\cite{Flam09}. There have also been several studies of the implications of Oklo 
for changes in $X_s=m_s/\Lambda$ ($m_s$ being the mass of the strange quark), but recent model independent
results \cite{BEFH13,Kron12} on pertinent hadronic properties (sigma terms) are discouraging. (In this subsection, 
all masses are in units such that $\Lambda=1$.)

Central to the approach of Flambaum and Wiringa is their conjecture that the non-Coulombic contribution 
to the any shift in $E_r$ is independent of mass number $A$ (and proton number $Z$). More precisely, 
their working assumption is that this property applies to the energies of any weakly bound states
as well as any resonances close to the neutron escape threshold. Arguments based on the Fermi gas model 
(ignoring configuration-mixing) and a small sample of quantitative results (for $A=6$ to $A=9$ nuclei) are given 
in support of this conjecture and then it is invoked to infer that
\begin{equation}\label{Xq}
  m_q\, \frac{dE_r}{dm_q} \sim 10\,\mathrm{MeV} 
\end{equation}
from calculations in light nuclei of the kind performed in Ref.~\cite{Flam07}. In fact,
Flambaum and Wiringa compute shifts not in $E_r$ \emph{per se\/} but in the difference of ground 
states energies
\[
 S \equiv \langle g|\widehat{H} |g\rangle - \langle \widetilde{g}|\widehat{H} |\widetilde{g}\rangle,
\]  
which is related to $E_r$ by $E_r=E^*-S$, $E^*$ being the excitation energy of the compound nucleus state $|r\rangle$
relative to the ground state $|\widetilde{g}\rangle$ of the same nucleus. (If $|r\rangle$ is an excited
state of the nucleus ${}^A_Z\mathrm{X}$, then $|g\rangle$ is the ground state of the isotope 
$\nucl{A-1}{Z}{X}$.)
It is tacitly assumed in Ref.~\cite{Flam09} that shifts in $E^*-S$ resemble those in $S$. The result in 
Eq.~(\ref{Xq}) is finally corroborated with an estimate based on a 
nuclear structure model appropriate to the study of heavy nuclei (the Walecka model).

\setlength{\tabcolsep}{3pt}

\begin{table}[pb] 
\caption{$K_E^H$ and $K_E$ values for $A=5$ to $A=9$ nuclei (AV18+UIX Hamiltonian).\footnote{Unless indicated 
otherwise, $K_E^H$ values are from Table IV in Ref.~\cite{Flam07}. Sets 2a, b and c of $K_E$'s are obtained with 
set 2 of the $K_H^q$'s in Table \ref{tb:Kq} and $K_V^q=0.6$, 0.7 and 0.8, respectively (choices motivated in the 
text).}\label{tb:KH}}
{\begin{tabular}{@{}l*{10}{c}@{}} \toprule
$H$     & ${}^5$He                    & ${}^5$Li                     & ${}^6$He           & ${}^6$Li           & ${}^6$Be           & ${}^7$He               & ${}^7$Li           & ${}^7$Be           & ${}^8$Be           & ${}^9$Be                \\ \colrule
$N$     &\hphantom{$-$}13.31 &\hphantom{$-$}13.66\footnotemark[4]&\hphantom{$-$}15.78\footnotemark[4]& \hphantom{$-$}14.41&\hphantom{$-$}17.30\footnotemark[4] & \hphantom{$-$}19.34\footnotemark[2]& \hphantom{$-$}15.53& \hphantom{$-$}16.29& \hphantom{$-$}14.36& \hphantom{$-$}16.09\footnotemark[2] \\ 
$\Delta$& $-$10.24                 & $-$10.54\footnotemark[4]                   
& $-$12.25\footnotemark[4]                 & $-$10.80                   & $-$13.48\footnotemark[4]                  
& $-$14.92\footnotemark[2]           & $-$11.96           & $-$12.56           & $-$11.11           
& $-$12.39\footnotemark[2]            \\ 
$\pi$   & \hphantom{0}$-$5.82&\hphantom{0}$-$5.98\footnotemark[4]&\hphantom{0}$-$7.02\footnotemark[4]
& \hphantom{0}$-$6.31& \hphantom{0}$-$7.72\footnotemark[4]& \hphantom{0}$-$8.78\footnotemark[2]& \hphantom{0}$-$6.91
& \hphantom{0}$-$7.26& \hphantom{0}$-$6.31& \hphantom{0}$-$7.27\footnotemark[2] \\ 
$V$     & \hphantom{$-$}40.87&\hphantom{$-$}42.04\footnotemark[4] &\hphantom{$-$}49.09\footnotemark[4]
& \hphantom{$-$}43.48&\hphantom{$-$}54.04\footnotemark[4]& \hphantom{$-$}60.46\footnotemark[2]& \hphantom{$-$}48.11
& \hphantom{$-$}50.53& \hphantom{$-$}44.40& \hphantom{$-$}50.21\footnotemark[2] \\ \colrule
$K_E$ & & & & & & & & & &  \\
set $1$\footnotemark[3]           
        &$-1.24$\hphantom{0} &  $-1.28$\hphantom{0}\footnotemark[4]     & $-1.50$\hphantom{0}  &$-1.36$\hphantom{0} & $-1.67$\hphantom{0}\footnotemark[4] & $-1.93$\hphantom{0}    &$-1.50$\hphantom{0} & $-1.57$\hphantom{0}& $-1.35$\hphantom{0}& $-1.59$\hphantom{0}     \\ 
set 2a
        &\hphantom{$-$}0.011 &\hphantom{$-$}0.013\hphantom{${}^c$}  &    $-0.010$                &$-0.033$            & $-$0.011\hphantom{${}^c$} &$-0.080$                &$-0.021$            & $-0.024$           & \hphantom{$-$}0.014& $-0.054$                 \\ 
set 2b
        &\hphantom{$-$}0.42\hphantom{0}& \hphantom{$-$}0.43\hphantom{$0^c$}  &\hphantom{$-$}0.48\hphantom{0} &\hphantom{$-$}0.40\hphantom{0}& \hphantom{$-$}0.53\hphantom{$0^c$}     &\hphantom{$-$}0.52\hphantom{0} &\hphantom{$-$}0.46\hphantom{0}&\hphantom{$-$}0.48\hphantom{0}& \hphantom{$-$}0.46\hphantom{0}&\hphantom{$-$}0.45\hphantom{0} \\ 
set 2c
        &\hphantom{$-$}0.83\hphantom{0}&\hphantom{$-$}0.85\hphantom{$0^c$} & \hphantom{$-$}0.97\hphantom{0} &\hphantom{$-$}0.84\hphantom{0}&\hphantom{$-$}1.07\hphantom{$0^c$}    &\hphantom{$-$}1.13\hphantom{0} &\hphantom{$-$}0.94\hphantom{0}&\hphantom{$-$}0.99\hphantom{0}& \hphantom{$-$}0.90\hphantom{0}&\hphantom{$-$}0.95\hphantom{0} \\ \botrule
\end{tabular}}
\footnotetext[2]{From Table III in Ref.~\cite{Flam09}.}
\footnotetext[3]{From Table I in Ref.~\cite{Flam09}.}
\footnotetext[4]{R. B. Wiringa, private communication.}
\end{table}

\setlength{\tabcolsep}{6pt}

To establish the quark mass dependence of $S$ for light nuclei, Flambaum and Wiringa employ the variational 
Monte Carlo method together with the Argonne $v_{18}$ (AV18) two-nucleon and Urbana IX (UIX) three-nucleon 
interactions \cite{Piep01}. They determine how ground state energies change for independent 0.1\% variations 
in the masses $m_H$ of the hadrons $H$ deemed most important to low-energy nuclear dynamics: nucleons ($H=N$), 
deltas ($H=\Delta$), pions ($H=\pi$) and, lastly, a vector meson ($H=V$), which simulates the short-range 
interactions associated with the $\rho$ and $\omega$ mesons.%
\footnote{Actually, the AV18+UIX Hamiltonian only depends explicitly on the masses of the pions, the 
proton and the neutron. Details of how the effects of changes in $m_\Delta$ and $m_V$ are found appear 
in Sec.~II.B of Ref.~\cite{Flam07}.} Their results are reported as the dimensionless response coefficients
\[
   K_E^H = \frac{m_H}{E}\, \frac{\delta E}{\delta m_{\mathrlap{H}}} \hspace*{1.4ex} ,
\]
where $E$ is the unperturbed ground state energy [in Refs.~\cite{Flam09} and \cite{Flam07}, $K_E^H$ is 
denoted by $\Delta{\cal E}(m_H)$]. The changes in $m_H$ are related to changes in $m_q$ by 
the sigma terms
\[
   \sigma_H = m_q\, \frac{d m_H}{dm_q} \equiv\, K_H^q\, m_H
\]
inferred from studies of hadronic structure. Thus, in terms of the ground state energies $E_g=\langle 
g|\widehat{H} |g\rangle$ and $E_{\widetilde{g}}=\langle\widetilde{g}|\widehat{H} |\widetilde{g}\rangle$,
\begin{equation}\label{sSmq}
 \widetilde{\sigma}_{\!\scriptscriptstyle S} \equiv m_q \frac{dS}{dm_q} = K_{E_g} E_g - K_{E_{\widetilde{g}}} E_{\widetilde{g}},
\end{equation}
where, via the chain rule for rates of change, the sensitivity 
\[
    K_E = \sum\limits_H K_E^H K_H^q  ,
\]
the sum being over the hadrons identified above. 

\begin{table}[pt]
\caption{$K_H^q$ values.\label{tb:Kq}}
{\begin{tabular}{@{}l*{6}{c}@{}} \toprule
 $H=$                                &$N$       & $\Delta$ & $\pi$    & $V$             & $\rho$   & $\omega$   \\ \colrule
Set 1 (used in Ref.~\cite{Flam09})&0.064\footnotemark[1] & 0.041\footnotemark[1]& 0.498\footnotemark[1]& 0.03\hphantom{0}& 0.021\footnotemark[2]& 0.034\footnotemark[2]  \\ 
Set 2 (more recent)                  &0.048\footnotemark[3] & 0.020\footnotemark[4]& 0.494\footnotemark[3]&                 
& 0.058\footnotemark[3]&            \\ \botrule
\end{tabular}}
\footnotetext[1]{From Ref.~\cite{FHJR06}, Eq.~(85).}
\footnotetext[2]{From Ref.~\cite{HMRW06}, Table 2.}
\footnotetext[3]{From Ref.~\cite{BEFH13}, Sec.~2.}
\footnotetext[4]{From the RL2 with pion exchange result for $\sigma_{\pi\Delta}$ in Eq.~(16) of Ref.~\cite{SAFK14}.}
\end{table}

Relevant values of $K_E^H$ and $K_H^q$ are given in Tables \ref{tb:KH} (upper half) and \ref{tb:Kq}, respectively. 
Barring $K_\pi^q$, recent values of the coefficients $K_H^q$ (Set 2) are appreciably different from 
those used in Ref.~\cite{Flam09} (Set 1). There is also the issue of what to adopt for $K_V^q$. 
The value chosen in Ref.~\cite{Flam09} is the average (to one significant figure) of  $K_\rho^q$ and 
$K_\omega^q$ in Set 1. We can make a similar estimate of $K_V^q$ with the value of $K_\rho^q$ in Set 2; we appeal 
to the fact that, in each of the two calculations \cite{FHJR06,HMRW06} known to us in which both $K_\rho^q$ 
and $K_\omega^q$ are determined, $K_\omega^q-K_\rho^q=0.013$: consequently, our preferred value of $K^q_V$ is
\[
  K^q_V = \textstyle{\frac{1}{2}} (2K_\rho^q + 0.013) = 0.06
\]
to one significant figure. In view of the uncertainty in $K^q_V$, we have generated three sets of
values of $K_E$ (sets 2a, 2b and 2c in Table \ref{tb:KH}) using set 2 of $K_H^q$'s in Table \ref{tb:Kq}
and $K_V^q=0.6$, 0.7 and 0.8, respectively. For comparison, the sensitivities quoted in
Ref.~\cite{Flam09} are included as set 1 in Table~\ref{tb:KH}.

The corresponding values of $\widetilde{\sigma}_{\!\scriptscriptstyle S}$ are presented in Table~\ref{tb:Sa}. 
The choice of Hamiltonian is such that ground state energies coincide with binding energies and we have used 
binding energies taken from experiment in evaluating Eq.~(\ref{sSmq}). In view of the scatter of values in Table \ref{tb:Sa}, no 
firm conclusions about the order of magnitude of $\widetilde{\sigma}_{\!\scriptscriptstyle S}$ are possible,
except that the estimate in Ref.~\cite{Flam09} of $|\widetilde{\sigma}_{\!\scriptscriptstyle S}|\sim 
10\,\mathrm{MeV}$ is an overestimate. In the case of set 2a in Table~\ref{tb:Sa} (our preferred choice), 
$|\widetilde{\sigma}_{\!\scriptscriptstyle S}|\lesssim 1\,\mathrm{MeV}$.

\begin{table}[pb]
\caption{$\widetilde{\sigma}_S$ in Eq.~(\ref{sSmq}) for some $A=5$ to $A=9$ nuclei (using experimental binding energies $E_g^\mathrm{exp}$).\label{tb:Sa}}
{\begin{tabular}{@{}l*{9}{c}@{}} \toprule
        & ${}^5$He         & ${}^6$He           & ${}^6$Li          & ${}^7$He               & ${}^7$Li            & ${}^7$Be           & ${}^8$Be           & ${}^9$Be                \\ \colrule
$E_g^\mathrm{exp}$     
        &  $-$27.41          & $-$29.27        & $-$31.99           & $-$28.83               & $-$39.24            & $-$37.60           & $-$56.50           & $-$58.16  \\ \colrule
$\widetilde{\sigma}_{\scriptscriptstyle S}\ $(set 1)\footnote{From Table IV in Ref.~\cite{Flam09} 
($\widetilde{\sigma}_{\!\scriptscriptstyle S}=m_q\,\delta S_\mathrm{expt}/\delta m_q$ in the notation of Ref.~\cite{Flam09}).}         
        &  $-3.42$           & $-9.92$          & $-9.52$          &  $-11.7$\hphantom{0}  &$-15.4$\hphantom{0}& $-15.5$\hphantom{0}          &$-17.2$\hphantom{0}& $-16.2$\hphantom{0} \\ 
\hphantom{$\sigma_{\scriptscriptstyle S}\ $}(set 2a)
        &  $-0.75$           & $-0.60$                &    $-1.39$ & \hphantom{0}$-2.01$             &\hphantom{$-0$}0.23& \hphantom{0}$-0.62$ &\hphantom{$-0$}1.67&\hphantom{0}$-$3.94 \\ 
\hphantom{$\sigma_{\scriptscriptstyle S}\ $}(set 2b)
        &\hphantom{$-$}$0.17$& \hphantom{$-$}2.57       &\hphantom{$-$}1.45 & \hphantom{0$-$}1.06          &\hphantom{$-0$}5.20&   \hphantom{0$-$}3.83    &\hphantom{$-0$}7.77&\hphantom{0$-$}0.17 \\ 
\hphantom{$\sigma_{\scriptscriptstyle S}\ $}(set 2c)
        &\hphantom{$-$}$1.09$& \hphantom{$-$}5.73      &\hphantom{$-$}4.29 &  \hphantom{0$-$}4.12  &\hphantom{$-$}10.2\hphantom{0}&  \hphantom{0$-$}8.29     &\hphantom{$-$}13.9\hphantom{0} &\hphantom{0$-$}4.29 \\ \botrule
\end{tabular}}
\end{table}

Unfortunately, as inspection of the $K_E^V$ coefficients in Table \ref{tb:KH} and the manner in which changes
in $m_V$ are effected (cf.~Sec.~II.B in Ref.~\cite{Flam07}) reveals,
$\widetilde{\sigma}_{\!\scriptscriptstyle S}$ is most sensitive to that part of the
AV18+UIX Hamiltonian which is least well connected to the properties of specific hadrons.
However, there exists a framework for systematically relating phenomenologically
successful nuclear forces to effective field theories appropriate to low-energy QCD \cite{EHM09,ME11,HNS13}.
A characterization of the short range part of the AV18 potential in terms of the low-energy constants 
(LECs) of an effective Lagrangian is known \cite{EMGE02}, and, recently, 
the quark mass dependence of LECs for the ${}^1S_0$ and ${}^3S_1-{}^3D_1$ partial waves has been established \cite{BEFH13}.
It is, perhaps, not too much to hope that some fruitful combination of these developments may 
circumvent the issues thrown up by the vector meson $V$. There remains, of course, the treatment of the
three-nucleon UIX potential but contributions to the $K_E^H$'s from its two-pion part 
have been found to be small \cite{Flam07}.

The Flambaum-Wiringa conjecture is an important idea, arguably a \emph{sine qua non\/} for
the reliable extraction of information on $X_q$ from Oklo data. More evidence in support of this conjecture 
is essential. Microscopic calculations for medium-heavy nuclei with the
renormalized Fermi hypernetted chain method \cite{ABC07} would be a challenging but helpful line of investigation.

\subsection{Unified treatment of the sensitivities to $\alpha$ and $X_q$}

On the basis of the results in the two previous subsections, we postulate the following relation for the shift
$\Delta_r\equiv E_r(\mathrm{Oklo}) - E_r(\mathrm{now})$ in the position of a resonance (near threshold) due to 
(small) changes $\Delta X_q\equiv X_q(\mathrm{Oklo}) - X_q(\mathrm{now})$ and $\Delta\alpha\equiv \alpha(
\mathrm{Oklo}) - \alpha(\mathrm{now})$ in $X_q$ and $\alpha$, respectively:
\begin{equation}\label{glb}
  \Delta_r  = a\, \frac{\Delta X_q}{X_q} + b \frac{Z^2}{A^{4/3}}\, \frac{\Delta \alpha}{\alpha} ,
\end{equation}
where the coefficients $a$ and $b$ are independent of $A$ and $Z$. The lack of any dependence on $A$ 
and $Z$ in the first term is conditional on the validity of the Flambaum-Wiringa conjecture.
The scaling with $A$ and $Z$ of the second term is, in part, deduced from Eq.~(\ref{ome}) on substitution of the
mass number dependence of $\left[r^2\right]_{\widetilde{g}g}$ implied by the \emph{uniform shift\/} formula
(cf. Eq.~(48) in Ref.~\cite{Ott89}). We also have to assume that the integral discarded in Eq.~(\ref{myVrint}) 
to obtain Eq.~(\ref{ome}) shares this scaling.

Our deliberations in subsection \ref{sc:alpha} suggest it is
reasonable to assume that $|b|\sim 1\,\mathrm{MeV}$. The order of magnitude of $a$ is less
certain: according to Ref.~\cite{Flam09}, $a\sim 10\,\mathrm{MeV}$, but the results
in Table \ref{tb:Sa} indicate that $a$ could be one or even two orders of magnitude smaller
($a$ is the average of $-\widetilde{\sigma}_{\!\scriptscriptstyle S}$ for a given set of $K_H^q$'s).

With a large enough data set, the different dependences on mass and proton number in Eq.~(\ref{glb}) should permit
one to disentangle the contributions of $\Delta X_q$ and $\Delta\alpha$ without any assumptions about their
relative size. Independent limits on $\Delta X_q$ and $\Delta\alpha$ would open up the possibility of
constraining the mechanism for time variation along the lines pursued 
in Refs.~\cite{Cal02b}, \cite{Frit09} and \cite{Frit11}. Apart from Ref.~\cite{SV91} (which 
invokes a questionable ``mean scaling'' hypothesis), we do not know of any Oklo 
studies which involve more than a couple of nuclei.

An entirely model independent limit on either $\Delta X_q$ or $\Delta\alpha$ is not possible for a \emph{single\/} 
nucleus, unless one of the two terms in Eq.~(\ref{glb}) is known \emph{a priori\/} to be dominant. In the case of 
$^{149}$Sm, it has been customary to discard the $X_q$-term, a step which has been strongly criticized on 
two grounds. First, there are studies (notably, Ref.~\cite{Flam09}) which find that the coefficient $a$ of 
the $X_q$-term is an order of magnitude larger than the coefficient multiplying $\Delta\alpha/\alpha$; second,
calculations \cite{Lan02,Cal02} based on the Callen-Symanzik renormalization group equation suggest that, within 
any theory which admits the unification of the Standard Model gauge couplings
(at a mass scale $\Lambda_u$ below the new physics responsible for the time variation of fundamental constants),
the evolution of these couplings to lower energies is such that $\left|\Delta  X_q/X_q\right|$ is an order 
of magnitude larger than $\left|\Delta \alpha/\alpha\right|$ if the behavior of other Standard Model 
parameters is ignored (and $\Lambda_u$ is not 
time-dependent). These objections do not stand up to closer inspection.

Our results on $a$ and $b$ indicate that the coefficients of $\Delta  X_q/X_q$ and 
$\Delta \alpha/\alpha$ in Eq.~(\ref{glb}) are more likely to be comparable. As regards the relative magnitudes of 
$\Delta X_q/X_q$ and $\Delta \alpha/\alpha$, the most complete statement implied by 
the analysis of Ref.~\cite{Lan02} is (cf.~Eq.~(30) in Ref.~\cite{Lan02})
\begin{equation}\label{Xal}
 \Bigl| \frac{\Delta X_q}{X_q}  \Bigr| \sim \Big| (R - \lambda - 0.8\kappa )\, \frac{\Delta \alpha }{\alpha} \Bigr| , 
\end{equation}
where $R\simeq\pi/[9\alpha(M_{\!\scriptscriptstyle Z})]$, $\alpha(M_{\!\scriptscriptstyle Z})$ being the electromagnetic 
coupling at the electroweak scale $M_{\!\scriptscriptstyle Z}$ [$\alpha(M_{\!\scriptscriptstyle Z})^{-1}=127.9$],
and $\lambda$ and $\kappa$ parametrize the time-dependence of Yukawa couplings to fermions and the vacuum expectation 
value of the Higgs boson, respectively.%
\footnote{Cf.~Eqs.~(11) and (14), respectively, in Ref.~\cite{Lan02}. More precisely, $\lambda$ denotes the average 
value of the $\lambda_a$'s in Eq.~(11) of Ref.~\cite{Lan02}.}
Equation (\ref{Xal}) should be compared with the result in Ref.~\cite{Lan02} for the variation in 
$\mu\equiv m_p/m_e$, namely 
\begin{equation}\label{murel}
  \frac{\Delta  \mu }{\mu} \sim (R - \lambda - 0.8\kappa )\, \frac{\Delta\alpha}{\alpha}  
\end{equation}
[cf.~Eq.~(20) in Ref.~\cite{Lan02}, where $Y$ is used instead of the more standard notation $\mu$], and
the experimental results
\begin{equation}\label{dela}
     \frac{\Delta  \alpha}{\alpha} = (-7.4\pm 1.7)\times 10^{-6}  
\end{equation}
and
\begin{equation}\label{delmu}
    \frac{\Delta \mu }{\mu} = (2.6\pm 3.0)\times 10^{-6}
\end{equation}
for overlapping red shifts $1.8 < z < 4.2$ and $z\sim 2.81$, respectively \cite{MFWD04,KWMC08}.
If the expression for $\Delta\mu/\mu$ obtained on combining Eqs.~(\ref{murel}) and (\ref{dela})
is to be compatible with the bound in Eq.~(\ref{delmu}), then the value of the (constant) factor $R-\lambda-0.8\kappa$ 
must be such that $\left|\Delta X_q/X_q\right| \sim \left|\Delta \alpha/\alpha\right|$. 
Thus, under circumstances in which the rather general model of Ref.~\cite{Lan02} applies,
it is permissible to ignore the $\Delta X_q/X_q$-term in Eq.~(\ref{glb})
when extracting an order of magnitude limit on $\Delta\alpha/\alpha$, and \emph{vice versa}.

\begin{figure}[bp]
\centering
\includegraphics[width=7.6cm]{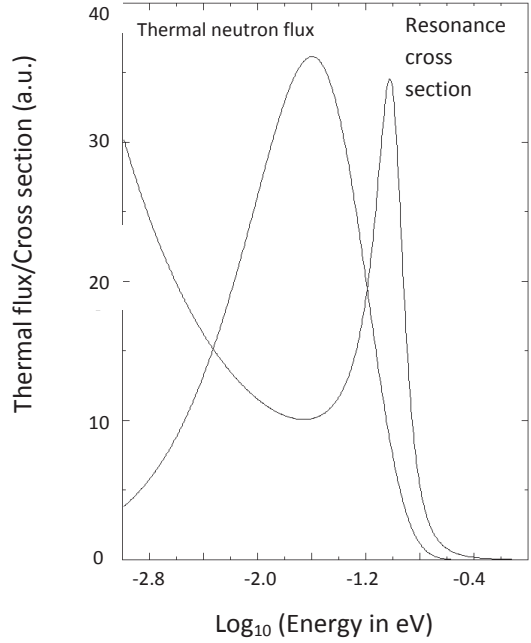}
\caption{Overlap of the 97.3 meV resonance of ${}^{149}$Sm  with a thermal neutron spectrum. If the energy of the 
resonance is different now compared to 2 Gyr ago, burn up of ${}^{149}$Sm is changed and the isotopic abundances 
remaining today will in principle indicate the magnitude and sign of the energy shift.}
\label{fig:thermalres}
\end{figure}

\subsection{Bound on the energy shift $\Delta_r$ of the Sm resonance}

The experimental basis for extracting a bound on the energy shift is illustrated in Fig.~\ref{fig:thermalres} which 
shows the overlap of a simple Breit Wigner resonance with a thermal neutron spectrum. If the energy of the Sm 
resonance shifts down from its present day value of  $E_r(\mathrm{now})=97.3\,\mathrm{meV}$, then more ${}^{149}$Sm 
will be burned in the neutron flux, and less ${}^{149}$Sm will be found in the isotopic remains of the reactor. 
Conversely, if the resonance shifts up, less will be burned, and the remains will be richer in ${}^{149}$Sm. The 
result is parametrized by an effective capture cross section $\hat{\sigma}$ which will depend on the amount the 
resonance is shifted. 

\begin{figure}[bp]
\centering
\includegraphics[width=8.6cm]{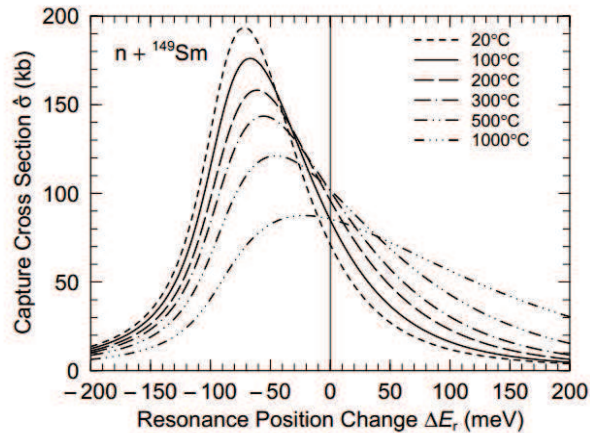}
\caption{Effective cross sections for ${}^{149}$Sm as a function of the 97.3 meV resonance shift, calculated with 
thermal neutron spectra of various temperatures. From Y. Fujii \emph{et al.},  Nucl. Phys. {\bf B573}  (2000) 377, 
with permission from Elsevier.} 
\label{fig:fujiism}
\end{figure}

\begin{figure}[tp]
\centering
\includegraphics[width=7.6cm]{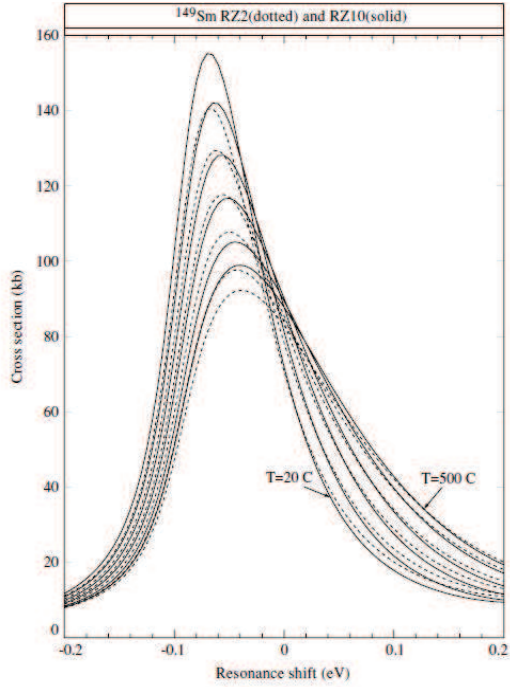}
\caption{Effective cross sections for ${}^{149}$Sm as a function of the 97.3 meV resonance shift, calculated with 
neutron spectra that include both thermal and epithermal components. The shapes are similar to those derived from 
thermal only calculations (see earlier), but differ in magnitudes.  From C. R.  Gould, E. I.  Sharapov and 
S. K. Lamoreaux, Phys. Rev. C {\bf 74} (2006) 24607.}
\label{fig:gouldsm}
\end{figure}

In practice, the temperature of the neutron spectrum must be taken into account, other resonances may have to be 
included, and details of the results can change when the epithermal component of the flux is included. This is 
illustrated in Figs.~\ref{fig:fujiism} and \ref{fig:gouldsm}  which show $\hat{\sigma}$ values from  
Refs.~\cite{FIFO00} and ~\cite{Gould1}, respectively. The former are calculated with thermal fluxes of 
different temperatures, the latter using realistic fluxes. While the shapes are similar, the differences in 
magnitudes are relevant when it comes to comparing with experimentally derived $\hat{\sigma}$ values. 

The cross sections are obtained by solving coupled equations \cite{Gould1,FIFO00} which seek to reproduce the 
isotopic abundancies reported for samples from the reactor zones \cite{Hid98}. The equations take into account 
isotope production in fission, generation of plutonium through neutron capture, and isotope burn up in the neutron 
flux. Post processing contamination is also sometimes included as an additional parameter in analyses \cite{FIFO00}. 

As is to be expected from geological samples, while trends of isotopic depletion and change are clear, variations 
outside of the statistical uncertainties are seen and further complicate the extraction of a resonance shift. In 
Ref.~\cite{Gould1} this was taken into account for RZ10 by not analyzing each sample individually but, instead, 
analyzing a meta sample, the average of the isotopic data for the four samples. 

\begin{figure}[tp]
\centering
\includegraphics[width=7.6cm]{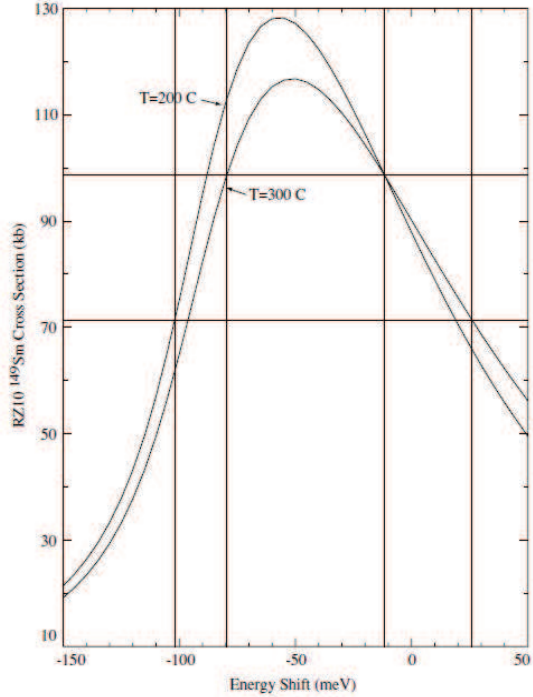}
\caption{Effective cross sections for ${}^{149}$Sm at  200~$^\circ$C and 300~$^\circ$C, and bounds (vertical and 
horizontal lines) indicating allowed solutions for the energy shift based on isotopic abundance data.  From C. R.  
Gould, E. I.  Sharapov and S. K. Lamoreaux, Phys.\ Rev.\ C {\bf 74} (2006) 24607.}
\label{fig:limits}
\end{figure}

The effective RZ10 neutron capture cross section for ${}^{149}$Sm determined in this way was  
$(85.0 \pm 6.8)\,\mathrm{kb}$ and, as seen in Fig.~\ref{fig:limits}, this leads to two solutions for the energy 
shift $\Delta_r$: a right branch overlapping zero, 
\begin{equation}\label{Delbd}
-11.6\,\mathrm{meV} \le \Delta_r \le 26.0\,\mathrm{meV}, 
\end{equation}
and a left branch yielding a non-zero solution, $-101.9\,\mathrm{meV} \le \Delta_r \le -79.6\,\mathrm{meV}$.

Similar double-valued solutions were found in Ref.~\cite{FIFO00}, where gadolinium isotopic data were used to 
try and establish one or other of the solutions as more probable. An implication of Eq.~(\ref{glb}) is that 
$\Delta_r$ should be the same to within a percent of so for the Sm and Gd data. Without the benefit of 
Eq.~(\ref{glb}), the authors of Ref.~\cite{FIFO00} had to make the seemingly \emph{ad hoc\/} assumption that
Sm and Gd should give the same result for $\Delta_r$. The analysis of 
Ref.~\cite{FIFO00} favored the zero shift solution but 
was complicated by the post processing contamination issues mentioned earlier. At this time, while nearly all 
analyses are consistent with a zero shift, the non-zero solution of the left-hand branch is not ruled out on 
experimental grounds. 

Perhaps, Eq.~(\ref{glb}) can be the basis for a strategy to decide conclusively on the interpretation of the
left-hand branch of solutions.

\subsection{Limit on $\Delta\alpha$ and  $\Delta X_q$ implied by bound on $\Delta_r$ for the Sm resonance}

The data on root-mean-square charge radii in Table XII of Ref.~\cite{FBH95} implies that the isotopes 
${}^{149}$Sm and ${}^{150}$Sm have \emph{equivalent\/} charge radii of $6.4786(10)\,\mathrm{fm}$ and
$6.5039(12)\,\mathrm{fm}$ in their ground states $|g\rangle$ and $|\widetilde{g}\rangle$, respectively. 
(The more recent but less precise data in Table 6 of 
Ref.~\cite{LRS07} are compatible with these results.) As the ${}^{150}$Sm compound nucleus state $|r\rangle$ 
is just above the neutron escape threshold, the excitation energy is about 
$0.4\,\mathrm{MeV}$  per valence nucleon. This fact, in conjunction with the subshell spacing (in the vicinity 
of the Fermi levels) of the single particle level schemes for ${}^{150}$Sm (see, 
for example, Figs.~4 and 5 in Ref.~\cite{SAD01}), leads us to conclude that the charge distribution
of $|r\rangle$ will not be significantly different from that of the ground state.
Certainly, we do not anticipate a 25\% increase in the equivalent charge radius to $8.11\,\mathrm{fm}$ (the 
value adopted in Ref.~\cite{Dam96}). Instead, we set $R_r=6.5\,\mathrm{fm}$, i.e.\ the value of the 
equivalent charge radius for the ${}^{150}$Sm ground state. 
According to the samarium data in Table X of the compilation in Ref.~\cite{FBH95} (which supersedes the 
data of Ref.~\cite{BSS80} used by Dyson and Damour),
\[
  [r^2]_{\widetilde{g}g} = 0.250 \pm 0.020\, \mathrm{fm}^2 .
\]
Equation (\ref{ome}) then evaluates to
(with $Z=62$)
\[
  \left| \alpha \frac{dE_r}{d\alpha} \right| \sim 2.5\,\mathrm{MeV} .
\]
Despite the substantially smaller choice of $R_r$ (which is the smallest physically acceptable one),
this revised estimate of $|\alpha\, dE_r/d\alpha|$ is still of the same order of magnitude as the lower bound 
on $\alpha\, dE_r/d\alpha$ used in Ref.~\cite{Dam96}.

Taking into account that Eq.~(\ref{ome}) is an overestimate [because of the omission of the cancellation 
discussed in connection with Eq.~(\ref{myVrint})], we advocate the use of the relation
\begin{equation}\label{fe}
 \left| \alpha \frac{dE_r}{d\alpha} \right| \sim 1\,\mathrm{MeV}
\end{equation}
in the analysis of the ${}^{149}$Sm data. In effect, Eq.~(\ref{fe}) differs from the standard result of 
Ref.~\cite{Dam96} only in that there is no longer any attempt to attach a confidence level.

Equation (\ref{fe}), together with the bound on $\Delta_r$ in Eq.~(\ref{Delbd}) and the relation
$\Delta_r\simeq \left( \alpha dE_r/d\alpha\right) \left(\Delta\alpha/\alpha\right)$, implies the bound
\begin{equation}\label{falbd}
   \left| \frac{\Delta\alpha}{\alpha}\right| < 3\times 10^{-8} .
\end{equation}
If the coefficient $a$ in Eq.~(\ref{glb}) is of the order of 1 MeV (the case for  set 2a of sensitivities $K_E$ in 
Table~\ref{tb:KH}), then a similar bound applies to $\Delta X_q/X_q$.

\begin{table}[pb]
\caption{Bounds on $\Delta\alpha/\alpha\equiv[\alpha(\mathrm{Oklo}) - \alpha(\mathrm{now})]/\alpha(\mathrm{now})$ from
the Sm resonance shift.\label{tb:alphabds}}
{\begin{tabular}{@{}*{5}{c}@{}} \toprule
 Ref.   & Zones      &Neutron spectrum  &$\Delta\alpha/\alpha\ (10^{-8})$
                                        &$\dot{\alpha}/\alpha\ (10^{-17}\,\mathrm{yr}^{-1})$\footnotemark[1]\\ \colrule
\cite{Dam96}
        & 2,5       & Maxwell 
                        & $-9\mapsto 11$                          & $-5.5\mapsto 4.5$                   \\ 
        &             & ($180-700\,{}^\circ\mathrm{C}$)  &                                  &                \\  \colrule
 \cite{FIFO02}
        & 10,13  & Maxwell                                       
                        & $-2\mapsto 0.2$                        & $-0.1\mapsto 1$                   \\
        &             & ($200-400\,{}^\circ\mathrm{C}$)  &                                  &                \\  \colrule
\cite{Petr06} 
        &   2        & MCNP4C\footnotemark[2]   
                        &  $-5.6\mapsto 6.6$                    &  $-3.3\mapsto 2.8$             \\                         
        &             & (Fresh core)          &             &            \\  \colrule
\cite{Gould1}
        &  2,10   & MCNP4C
                       & $-1.1\mapsto 2.4$         &  $-1.2\mapsto 0.6$           \\                         
        &             &  (Spectral indices)\footnotemark[3]   &                                 &            \\  \colrule
\cite{Oneg012}\footnotemark[4]
        &  3,5     & MCNP4C
                       & $-1.0\mapsto 0.7$          & $-0.4\mapsto 0.5$           \\                         
        &             & (Realistic fuel burn-up)             &             &            \\          \botrule
\end{tabular}}
\footnotetext[1]{Limits on the {\em{average\/}} rate of change of $\alpha$ over the time since the Oklo reactors ceased
(relative to the current value of $\alpha$). We take the age of the natural reactors to be 2 billion years.}
\footnotetext[2]{Spectrum of neutrons calculated with the code documented in Ref.~\cite{Brie00}.}
\footnotetext[3]{Model for spectrum consistent with measured Oklo epithermal spectral indices.}
\footnotetext[4]{The inequalities in Eq.~(9) of Ref.~\cite{Oneg012} need to be reversed.}
\end{table}

Since the publication of Ref.~\cite{Dam96}, it has been common practice to use the relation
\begin{equation}\label{trad}
         \frac{\Delta\alpha}{\alpha} = - \frac{\Delta_r}{M}
\end{equation}
with $M=1.1\,\mathrm{MeV}$
to infer a bound on $\Delta\alpha$ from a bound on $\Delta_r$. Section III.C of Ref.~\cite{Petr06} 
can be consulted for a comprehensive overview of results based on Eq.~(\ref{trad}) up to the publication 
of Ref.~\cite{Gould1}. Table \ref{tb:alphabds} below contains some features of this summary and updates it to 
include Refs.~\cite{Gould1} and \cite{Oneg012}. It should be clear from our reappraisal of 
Ref.~\cite{Dam96} (in subsection \ref{sc:alpha}) that we believe one should be a little circumspect 
about the way Eq.~(\ref{trad}) has been used in the past to restrict $\Delta\alpha$ and the average value of 
$\dot{\alpha}$ to intervals about zero. Order of magnitude estimates based on Eq.~(\ref{trad}) are, however, probably reliable.
Thus, we advocate reporting the result of, for example, Ref.~\cite{Gould1} as the bound in Eq.~(\ref{falbd}).
This bound is reduced by a factor of 3 in Ref.~\cite{Oneg012}.

Our guarded attitude towards Eq.~(\ref{trad}) is shared by the authors of Ref.~\cite{DSW09}. In assessing the 
limits on $\dot{\alpha}/\alpha$ of Refs.~\cite{Petr06} and \cite{Gould1} (in the last column of Table 
\ref{tb:alphabds}), they adopt the most conservative null bound of $|\dot{\alpha}/\alpha|\le 3\times 10^{-17}\, 
\mathrm{yr}^{-1}$ and argue that, by multiplying this bound by a factor of 3, they can compensate for the neglected 
effect of variations in $X_q$ (and any other parameters influencing nuclear forces). This factor of 3
is arbitrary (as pointed out in Ref.~\cite{DSW09}), but its use is taken for granted in subsequent 
studies \cite{Den10}. We do not understand the stated rationale for the factor of 3, but it can be viewed as an
\emph{ad hoc\/} way of accommodating partial cancellations between the $X_q$ and $\alpha$ contributions to $\Delta_r$.

\section{Conclusions}

Unravelling how the geosphere and the biosphere evolved together is one of the most fascinating tasks for modern science.  The Oklo natural nuclear reactors, basically formed by cyanobacteria two billion years ago, are yet another example of the surprises to be found in Earth's history. Since their discovery over forty years ago, the reactors have provided a rich source of information on topics as applied as can nuclear wastes be safely stored indefinitely to topics as esoteric as are the forces of physics changing as the Universe ages?
  
In this review, we have summarized nuclear physics interests in the Oklo phenomenon, focusing particularly on developments over the past two decades. Modeling the reactors has become increasingly sophisticated, employing Monte Carlo simulations with realistic geometries and materials which can generate both the thermal and epithermal fractions.  The water content and the temperatures of the reactors have been uncertain parameters.  We have discussed recent work pointing to lower temperatures than earlier assumed. Nuclear cross sections are input to all Oklo modeling and we have identified a parameter, relating to the capture by the $^{175}$Lu ground state of thermal neutrons, that warrants further investigation.  

The use of Oklo data to constrain changes in fundamental constants over the last 2 billion years has motivated much recent work. We have presented a critical reappraisal of the current situation, starting with the long-standing study of Damour and Dyson on sensitivity to the fine structure constant $\alpha$.
We conclude that their result can plausibly be used for order of magnitude estimates, but an 
investigation of how this conclusion may be affected by a more careful treatment of the Coulomb potential in the vicinity 
of the nuclear surface (and beyond) is warranted. The more recent analysis by Flambaum and Wiringa of the sensitivity to 
the average mass $m_q$ of the light quarks has been updated to incorporate the latest values of sigma terms. No firm 
conclusions about the reliability of Flambaum and Wiringa's estimate are possible (because of uncertainties 
surrounding the short-ranged part of the nuclear interaction), but it could be an overestimate by as much as 
an order of magnitude.

On the basis of the work in Refs.~\cite{Dam96} and \cite{Flam09}, we have suggested a formula for the 
unified treatment of sensitivities to $\alpha$ and $m_q$, namely Eq.~(\ref{glb}). It is an obvious synthesis, which 
has the advantage of making explicit the dependence on mass number and atomic number. We hope that it may prove useful in distinguishing between the contributions of $\alpha$ and $m_q$ in a model independent way or, at least, facilitating
an understanding of the significance of the non-zero shifts in resonance energies which have been found in 
some studies of Oklo data.

Appealing to recent data on variations in $\alpha$ and the proton-to-electron mass ratio $\mu$, we have demonstrated
that, within the very general model of Ref.~\cite{Lan02} (and contrary to widespread opinion), shifts in resonance energies 
are not any more sensitive to variations in $m_q$ than they are to variations in $\alpha$. When extracting an \emph{order of 
magnitude limit\/} on any change in $\alpha$, it is, thus, permissable to ignore any changes in $m_q$ and {\emph{vice versa}}
(provided the model of Ref.~\cite{Lan02} applies). In fact, 
we have argued that one can, at best, use null shifts to establish order of magnitude estimates of upper bounds.  
Bounds on $\Delta\alpha/\alpha$ have been presented. The most recent 
study \cite{Oneg012} of the Oklo data pertaining to the
 $97.3\,\mathrm{meV}$ resonance seen now in neutron capture by ${}^{149}$Sm 
implies that $|\Delta\alpha/\alpha|
\lesssim 1\times 10^{-8}$ (cf.~Table \ref{tb:alphabds} for more details).

\section*{Acknowledgements}

We would like to thank V. V. Flambaum and R. B. Wiringa
for responding to our queries about Ref.~\cite{Flam09}, and R. B. Wiringa for recalculating
some of the entries needed in Table~\ref{tb:KH}.
C. R. G. acknowledges support by the US Department of Energy, Office of Nuclear Physics, 
under Grant No.~DE-FG02-97ER41041 (NC State University).

\bibliography{oklo}

\end{document}